\documentstyle[aps,prb,multicol,epsfig]{revtex}
\begin{document}
\title{ Magnetic phases of the mixed-spin $J_1-J_2$  Heisenberg model 
on a square lattice}
\author{ N.B. Ivanov$^1$, J. Richter$^2$,  and D.J.J. Farnell$^3$}
\address{$^{1)}$ Max-Planck Institut f\"ur Physik Komplexer Systeme,  N\"othnitzer Str.  38,
D-01187 Dresden, Germany\cite{ivanov}\\
$^{2)}$  Institut f\"ur Theoretische Physik, Universit\"at Magdeburg,
PF 4120, D-39016 Magdeburg, Germany\\
$^{3)}$   School of Mechanical Engineering, University of Leeds, Woodhouse
Lane Leeds LS2 9JT, United Kingdom 
}
\date{\today}
\maketitle
\begin{abstract}
We study the zero-temperature phase diagram and the low-energy excitations
of a mixed-spin ($S_1>S_2$) $J_1-J_2$ Heisenberg model defined on a square lattice
by using a spin-wave analysis, the coupled cluster method, 
and the Lanczos exact-diagonalization technique. As a function of 
the frustration parameter $J_2/J_1$ ($ >0$),   the phase diagram 
exhibits  a quantized ferrimagnetic phase, a canted spin phase,  and
a  mixed-spin collinear phase.  The presented results point towards 
a strong disordering effect of the frustration 
and quantum spin fluctuations in the vicinity of the classical spin-flop
transition. In the extreme quantum system  $(S_1,S_2)=(1,\frac{1}{2})$,  we find
indications of  a new  quantum spin state  in the region $0.46< J_2/J_1<0.5$.
\end{abstract}
\draft
\begin{multicols}{2}
%%%%%%%%%%%%%%%%%%%%%%%%%%%%%%%%%%%%%%%%%%%%%%%%
\section{Introduction}
%%%%%%%%%%%%%%%%%%%%%%%%%%%%%%%%%%%
There has recently been an increasing interest 
in Heisenberg spin systems exhibiting quantum phases 
with an extensive  magnetic  moment. An intriguing example is
the bilayer Hall system at filling factor  $\nu =2$ which was 
shown to possess a canted spin  state
with spontaneously broken $U(1)$ symmetry.\cite{sarma,yang}   
A closely related phase diagram was studied
in the framework of  the bilayer quantum Heisenberg model  subject to
a perpendicular magnetic field.\cite{troyer}
Some specific properties of the quantum  phase transitions in
systems with quantum ferromagnetic phases
have been previously discussed  by using a special class of lattice   
models  with quantum-rotor degrees of freedom.\cite{sachdev}  
In the present paper  we study a  $J_1-J_2$  
Heisenberg model  which is a natural   (mixed-spin)  extention    
of  the well known     antiferromagnetic $J_1-J_2$  
model\cite{sushkov}  and exhibits most  of the
quantum magnetic phases found in the aforementioned rotor systems.

The model   is defined by  the Hamiltonian 
\begin{equation}\label{h}
{\cal H}=J_1\sum_{({\bf r,r'})}{{\bf S}_{1}}_{\bf r}\cdot {{\bf S}_{2}}_{\bf r'}
+J_2\sum_{[{\bf r,r'}]} 
\left({{\bf S}_{1}}_{\bf r}\cdot {{\bf S}_{1}}_{\bf r'}+ 
{{\bf S}_{2}}_{\bf r}\cdot {{\bf S}_{2}}_{\bf r'}\right) \, ,
\end{equation}
where $({\bf r,r'})$ and   $[{\bf r,r'}]$ denote pairs of 
nearest   and next-nearest  (diagonal)  sites of  the square lattice.
We choose  a checkerboard arrangement for  the spin
operators ${{\bf S}_{1}}_{\bf r}$ and  ${{\bf S}_{2}}_{\bf r}$ 
[${{\bf S}_{1}}_{\bf r}^2=S_1(S_1+1)$, ${{\bf S}_{2}}_{\bf r}^2=S_2(S_2+1)$,
and $S_1>S_2$] and introduce  the parameters   
$\alpha \equiv J_2/J_1$ ($J_1,J_2\geq 0$) and  $\sigma \equiv S_1/S_2$.   
With some minor   modifications, e.g.,   introducing a spatial anisotropy in 
the nearest-neighbor exchange  interaction,
the Hamiltonian (\ref{h}) could  describe a large class of  real mixed-spin 
compounds such as the  bimetallic molecular magnets.\cite{kahn} 

\begin{figure}
\centering\epsfig{file=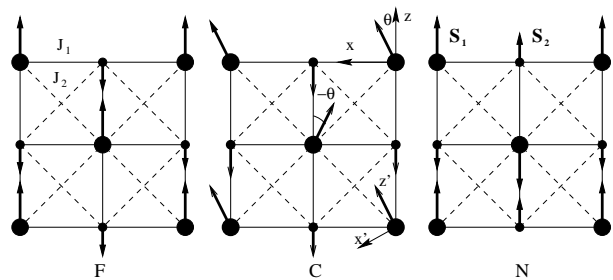,width=8cm}
\vspace{0.3cm}
\narrowtext
\caption{
The classical ferrimagnetic (F), canted (C), and  collinear (N)
spin phases of the mixed-spin $J_1-J_2$ model on a square lattice.
}
\label{phases}
\end{figure}

In the classical limit  the phase diagram of the mixed-spin system 
contains the  ferrimagnetic state (F phase),  the canted  
state (C phase), and   the  mixed-spin  collinear  state\cite{comm3} (N phase) 
shown in  Fig.~\ref{phases}. The latter phases are  stable, respectively,  
in the parameter regions
$\alpha < \alpha_{c1}$,  $\alpha_{c1}<\alpha < 0.5$, and
$\alpha >0.5$.  The classical F-C transition at 
$\alpha_{c1}=(2\sigma)^{-1}$ is continuous, whereas the C-N transition at 
$\alpha =0.5$ is connected with  a flop of the 
$S_1$ and $S_2$ spins. The  canted  spin phase  appears as  a result of the
magnetic  frustration without explicit  breaking of the $SU(2)$ symmetry
(in the mentioned bilayer systems the C phase is generated by a 
perpendicular magnetic field). Note also that the $S_2$ spins  
remain  collinear in the C phase. The angle $\theta$ measuring  the
local orientation of the classical $S_1$ spins in respect to the
global magnetization axis $z$  alternatively takes the values 
\begin{equation}
\theta=\pm \arccos \frac{1}{2\alpha\sigma}\, .
\end{equation}
Bimetallic molecular magnets are normally characterized  
by small exchange  constants $J_1$ such that the canted phase may 
appear  for   a moderate  magnetic frustration provided that
the parameter $\sigma$ is large enough.
In this article   we study   the quantum phase diagram of the model (\ref{h})
by using  the  standard
spin-wave theory (SWT),  the coupled cluster method\cite{farnell} (CCM), and
the Lanczos exact-diagonalization (ED) technique.
%%%%%%%%%%%%%%%%%%%%%%
%%%%%%%%%%%%%%%%%%%%%%%%%%%%%%%%%%%%%%%%%%%%%%%
%%%%%%%%%%%%%%%%%%%%%%
\section{Phase diagram}
\subsection{Spin-wave analysis}
%%%%%%%%%%%%%%%%%%%%%%%%%
To perform a SWT  analysis we  assume  
that the  classical spins  lie in the $xz$ plane. 
Using as a quantization  axis  the  local orientation 
of the classical spins (the $z'$ axis in Fig.~\ref{phases}), 
the  leading spin-wave terms in the expansions  
for  the spin operators ${{\bf S}_1}_{\bf r}$  and ${{\bf S}_{2}}_{\bf r}$ read  
\begin{eqnarray}
{S_{i}}_{\bf r}^z&=&\cos \theta_{i\bf r} (S_i-a_{i\bf r}^{\dag}a_{i\bf r})-\sin
\theta_{i\bf r}
\sqrt{\frac{S_i}{2}}(a_{i\bf r}^{\dag}+a_{i\bf  r}),
\nonumber\\
{S_{i}}_{\bf r}^x&=&\sin \theta_{i\_bf r}(S_i-a_{i\bf r}^{\dag}a_{i\bf r})+\cos
\theta_{i\bf r}
\sqrt{\frac{S_i}{2}}(a_{i\bf r}^{\dag}+a_{i\bf r}),
\nonumber\\
{S_{i}}_{\bf r}^y&=&\imath \sqrt{\frac{S_i}{2}}(a_{i\bf r}^{\dag}-a_{i\bf r})
\hspace{0.5cm} (i=1,2), \nonumber
\end{eqnarray}
where the angle $\theta_{i\bf r}$ measures the local deviations of $z'$  
from the magnetization axis $z$.  A substitution of  the latter    expressions   
in Eq. (\ref{h}) yields the following spin-wave  
Hamiltonian written   in terms of the Fourier  transforms $a_{i\bf k}$
of  the boson operators $a_{i\bf r}$:
\begin{eqnarray}\label{h0}
{\cal H}_{0}&=&\sum_{i=1}^2\sum_{\bf k}\left[ A_{i\bf k}a^{\dag}_{i \bf k}a_{i\bf
k}+\frac{B_{i\bf k}}{2}(a^{\dag}_{i\bf k}a^{\dag}_{i\bf -k}
+a_{i\bf k}a_{i\bf -k} )\right]\nonumber\\
&+&\sum_{\bf k}
\left[  C_{\bf k}a^{\dag}_{1\bf k}a^{\dag}_{2\bf -k}
+D_{\bf k}a^{\dag}_{1\bf k}a_{2\bf k}+h.c. \right] +E_0\, .
\end{eqnarray}
Note that  $E_0$ is the  energy  of the   
classical spin  state; the coefficients in Eq.~(\ref{h0}) are defined below.
The wave vector ${\bf k}=k_1{\bf e}_1+k_2{\bf e}_2$  runs
in the paramagnetic  Brillouin zone (PBZ) $|k_1\pm k_2|\leq \pi /a_0$ 
(the unit vectors ${\bf e}_1\perp {\bf e}_2$ point towards the 
main directions on the square lattice).
We shall henceforth use a system of units where the lattice spacing $a_0=1$. 

The  canonical transformation
$$
a_{i\bf k}=\sum_{j=1}^2\left[u^{(j)}_{i\bf k}\alpha_{j\bf k}+
v^{(j)}_{i\bf k}\alpha_{j-\bf k}^{\dag}\right]  \hspace{0.5cm} (i=1,2)   
$$
recasts  ${\cal H}_0$  to the diagonal form
\begin{equation}\label{h00}
{\cal H}_0=E_0-\sum_{\bf k}A_{\bf k}+
\sum_{i=1}^2\sum_{\bf k}\omega_i ({\bf k})\left( \alpha_{i\bf k}^{\dag}\alpha_{i\bf k}
+\frac{1}{2}\right) ,
\end{equation}
where  $A_{\bf k} =(A_{1\bf k}+A_{2\bf k})/2$.   
In the PBZ  any of the discussed magnetic phases is  characterized 
by two branches of spin-wave
excitations [$\omega_i=\omega_i ({\bf k})$, $i=1,2$] 
\begin{equation}\label{sp}
\omega_{1,2}=\sqrt{F_{\bf k}\mp\sqrt{G_{\bf k}}}\, ,
\end{equation}
where $F_{\bf k}=(\epsilon_{1\bf k}^2+\epsilon_{2\bf k}^2)/2+D_{\bf
k}^2-C_{\bf k}^2$, 
 $G_{\bf k}= (\epsilon_{1\bf k}^2-\epsilon_{2\bf k}^2)^2/4
+(\epsilon_{1\bf k}^2+\epsilon_{2\bf k}^2)(D_{\bf k}^2-C_{\bf k}^2)
+2(A_{1\bf k}A_{2\bf k}+B_{1\bf k}B_{2\bf k}) (C_{\bf k}^2+D_{\bf k}^2)
-4(A_{1\bf k}B_{2\bf k}+A_{2\bf k}B_{1\bf k})  C_{\bf k}D_{\bf k}$,
and $\epsilon_{i\bf k}=(A_{i\bf k}^2-B_{i\bf k}^2)^{1/2}$ ($i=1,2$).
$\omega_1$ and $\omega_2$
are the positive solutions of the  biquadratic algebraic equation 
\begin{equation}
\Delta (\omega)\equiv {\mathrm det}\: {\hat{\cal D }}(\omega)=0\, .
\end{equation}
The dynamical  matrix  $\hat{\cal D}(\omega)$ reads
$$
{ \hat{\cal D}}(\omega)=
\left[\begin{array}{cccc}
A_{1\bf k}-\omega  &  B_{1\bf k} & D_{\bf k} &  C_{\bf k} \\
B_{1\bf k}                 & A_{1\bf k}+\omega &C_{\bf k}&D_{\bf k} \\
D_{\bf k}&C_{\bf k}& A_{2\bf k}-\omega &B_{2\bf k} \\
C_{\bf k}&D_{\bf k}&B_{2\bf k}&A_{2\bf k}+\omega
\end{array} \right]\hspace{0.2cm}.
$$

A straitghforward  calculation gives the following expression 
for  the on-site magnetizations
$m_1\equiv \langle {S_1}_{\bf r}^{z'}\rangle$ and 
$m_2\equiv \langle {S_2}_{\bf r}^{z'}\rangle$: 
\begin{equation}\label{m}
m_i=S_i-\frac{2}{N}\sum_{\bf k}\left[
\left| v_{i\bf k}^{(1)}\right|^2
+\left| v_{i\bf k}^{(2)}\right|^2\right]\hspace{0.2cm} (i=1,2),
\end{equation}
where  
\begin{eqnarray}
 v_{1\bf k}^{(1)} &=&-\frac{\Delta_{12} (\omega_{1})}{\sqrt{
2\omega_{1}(\omega_2^2
-\omega_1^2)\Delta_{11}(\omega_1)}} \: ,\nonumber \\
 v_{2\bf k}^{(1)} &=& \frac{-\Delta_{14} (\omega_1)}{
\sqrt{2\omega_{1}(\omega_2^2
-\omega_1^2)\Delta_{11}(\omega_1)}} \: ,\nonumber \\
v_{1\bf k}^{(2)} &=& \sqrt{\frac{-\Delta_{11} (-\omega_2)}{
2\omega_2(\omega_2^2-\omega_1^2)}} \: ,\nonumber \\
v_{2\bf k}^{(2)} &=& \frac{-\Delta_{13} (-\omega_2)}{
\sqrt{-2\omega_2(\omega_2^2-\omega_1^2)\Delta_{11} (-\omega_2)}} \: .
\end{eqnarray}
In the above expressions, $\Delta_{ij}(\omega)$
is the minor of the $ij$ element of $\Delta (\omega)$ and $N$ is
the number of sites of the lattice.
%%%%%%%%%%%%%%%%%%%%%%
\begin{figure}
\centering\epsfig{file=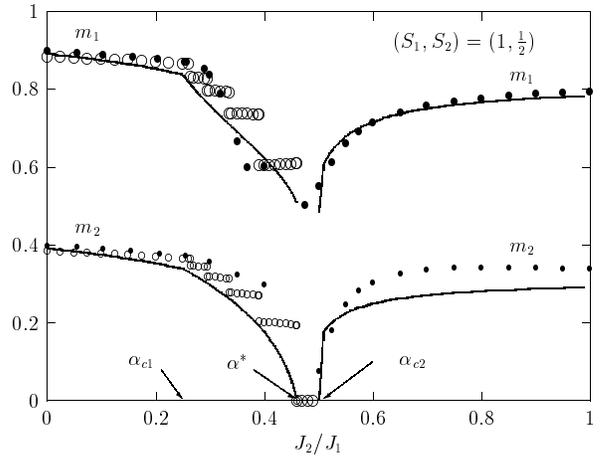,width=8cm}
\vspace{0.3cm}
\narrowtext
\caption{The on-site magnetizations $m_1= \langle {S_1}_{\bf r}^{z'}\rangle$
and $m_2=\langle {S_2}_{\bf r}^{z'}\rangle$ plotted as a function of
$J_2/J_1$:  SWT  (full lines), CCM
(filled circles), ED for $N=20$ spins 
and peroidic boundary conditions (open  circles). The ED data for $m_1$
in the C phase are obtained from the relation
$m_1=\langle {S_1}_{\bf r}^{z}\rangle/\cos \theta$ and the pitch angle
$\theta$ is extracted from the correlation function 
$K_{S_1}$ in Eq.~(\ref{ks}).
}
\label{mm}
\end{figure}
%%%%%%%%%%%%%%%%%%%%%%%%%%%%%%%%%%%%%%%%%%%%%
%%%%%%%%%%%%%%%%%%%%%%%%%%%%%%%%
\subsection{Magnetic phases}
%%%%%%%%%%%%%%%%%%%%%%%%%%%%%%%%

\begin{figure}
\centering\epsfig{file=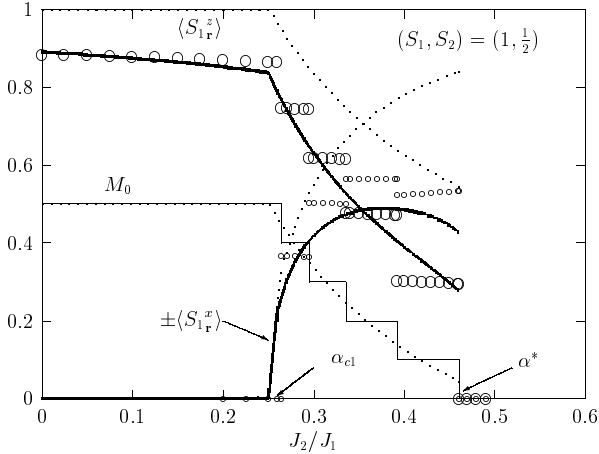,width=8cm}
\vspace{0.3cm}
\narrowtext
\caption{Order parameters in the F and C phases as obtained from the SWT (thick lines)
and the ED method for $N=20$ with periodic boundary conditions (open circles).
The thin broken line  shows the ED data for the magnetization per cell $M_0$ ($N=20$). 
The classical values of the parameters are shown by
dotted lines.
}
\label{mxmz}
\end{figure}

The spin-wave results for the  on-site magnetizations $m_1$ and 
$m_2$, Eq.~(\ref{m}),   in the   extreme quantum system   $(1,\frac{1}{2})$ 
are presented  in Fig.~\ref{mm}. We suppose that in the ferrimagnetic phase  
the sublattice magnetizations are oriented along the global $z$ axis,
i.e., $m_1=\langle {S_1}_{\bf r}^z\rangle$ and $m_2=-
\langle {S_2}_{\bf r}^z\rangle$. The F phase 
is also characterized by the  net 
ferromagnetic moment per cell $M_0=(S_1-S_2)=\frac{1}{2}$ 
(see Fig.~\ref{mxmz}). 
In the general case it is straightforward to observe that $M_0$ takes  only integer 
or half-integer values in the F phase. Following the authors of  
Ref.\onlinecite{sachdev},  this state  may  be reffered to as a  
{\em quantized unsaturated ferromagnetic phase}.  
The ED data from Figs.~\ref{mm} and \ref{mxmz} show that  the position 
of the continuous phase transition 
$\alpha_{c1}$  is slightly changed   by quantum fluctuations
towards larger $\alpha$.  $\alpha_{c1}$ is fixed as a point
where the spin of the ground state changes from $(S_1-S_2)N/2$
to $(S_1-S_2)N/2-1$. Respectively, in the $N=20$ system with
$(S_1,S_2)=(1,\frac{1}{2})$ the magnetization per cell changes from
$M_0=0.5$ to $M_0=0.4$ (Fig.~\ref{mxmz}). 
%%%%%%%%%%%%%%%%%%%%%%
%%%%%%%%%%%%%%%%%%%%%%%%%%%%%%%%%%%%%%%%%%%%%

\begin{figure}
\centering\epsfig{file=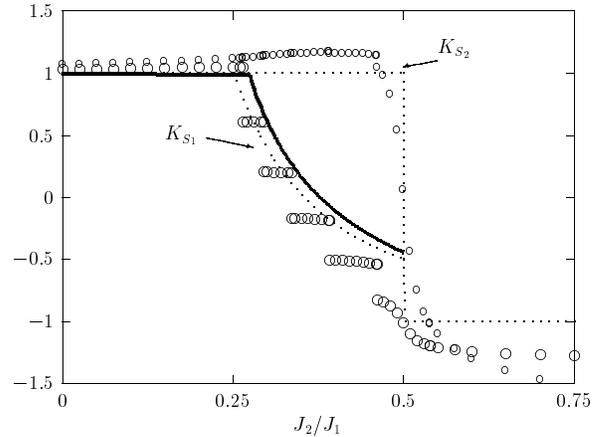,width=8cm}
\vspace{0.3cm}
\narrowtext
\caption{The spin-spin correlators $K_{S_1}$ and   $K_{S_2} 
$vs. $J_2/J_1$  used to obtain the pitch angle $\theta$
between neighboring spins in the $S_1$ and $S_2$ subsystems.  
The ED results ($N=20$) are represented 
by open circles. The dotted lines show the respective classical values.
The thick line represents the CCM results for  $\cos 2\theta$ in the 
SUB4-4  approximation.
}
\label{angle}
\end{figure}

More important changes take place  in the  C phase. The canted spin 
state is additionally  characterized  
by  the  staggered transverse field   
$\langle {S_1}_{\bf r}^x\rangle$  which 
breaks the spin rotation  symmetry $U(1)$ in the $xy$ plane.
In Fig.~\ref{mxmz} we present result for the transverse order parameter as
obtained from the SWT and the ED method. The ED data are obtained 
from the ED results for  $\langle {S_1}_{\bf r}^z\rangle$, by using 
the relation $\langle {S_1}_{\bf r}^x\rangle =
\langle {S_1}_{\bf r}^z\rangle \tan \theta$.
The pitch angle  $\theta$ is  extracted from 
the spin-spin correlation function (see Fig.~\ref{angle})\cite{wald} 
\begin{equation}\label{ks}
K_{S_1}=\frac{\langle {{\bf S}_1}_{\bf 0}\cdot {{\bf S}_1}_{\bf e_1+e_2}\rangle}{
\langle {{\bf S}_1}_{\bf 0}\cdot {{\bf S}_1}_{\bf 2e_1}\rangle}
\end{equation}
%%%%%%%%%%%%%%%%%%%%%%
%%%%%%%%%%%%%%%%%%%%%%%%%%%%%%%%%%%%%%%%%%%%%%%
which reduces to $K_{S_1}=\cos 2\theta$ in the classical limit $S_1\rightarrow \infty$.
 The respective correlation function  $K_{S_2}$ for the $S_2$ subsystem can be
  defined in  a similar way.   
 Figs.~\ref{mm} and \ref{mxmz} show that in the C phase the combined disordering effect
of the magnetic frustration and  quantum spin fluctuations
is strongly enhanced as  compared to   the F phase.  
SWT predicts a complete magnetic disordering of the $S_2$ 
spins  ($m_2=0$) starting  at $\alpha=\alpha^*$.
The latter point  exists in the SWT
for arbitrary spins $S_1$ and $S_2$\cite{comm0}
and preceeds the classical spin-flop transition
point ($\alpha =0.5$).
The   effect is strongly pronounced in the extreme quantum case
$(1,\frac{1}{2})$ where  both the SWT and the ED data 
predict  $\alpha^*\approx 0.46$. The extrapolated CCM data
seem to show the same tendency. 
The steps in the ED data  
are due to finite-size effects
and reflect   the change in the total spin $S_t$  (a good quantum number) 
of the  absolute ground state  with the frustration parameter $J_2/J_1$. 
For instance, in the ED data $\alpha^*$ appears as a  point
where $S_t$ changes from  $S_t=1$ to $S_t=0$ (respectively, 
$M_0$ changes from $0.1$ to $0$). Although 
SWT formally predicts $m_1 >0$ even beyond $\alpha^*$,
the latter region  is not accessible  (for the standard SWT) 
as  the phase with $m_1>0$ and $m_2 =0$ does not appear in the
classical phase diagram.\cite{comm1}  
We suggest  that the point $\alpha^*$ is  
just the new position of the  quantum spin-flop transition (see below).
It is interesting to note that quantum fluctuations in the C phase
do not change substantially the parameters $M_0$ and $\theta$ 
(see Figs.~\ref{mxmz} and \ref{angle}). 

Finally, the mixed-spin collinear state has properties which are
similar to those of the collinear phase  in the antiferromagnetic 
$J_1-J_2$ model, i.e.,   it is a N\'eel-type magnetic state 
characterized by the on-site magnetizations $m_1,m_2\neq 0$. 
Both the SWT and the  CCM  predict that  the on-site magnetization
$m_2$  vanishes at a  point ($\alpha_{c2}$) which is  very close to 
the classical spin-flop transition point.
%%%%%%%%%%%%%%%%%%%%%%
\begin{figure}
\centering\epsfig{file=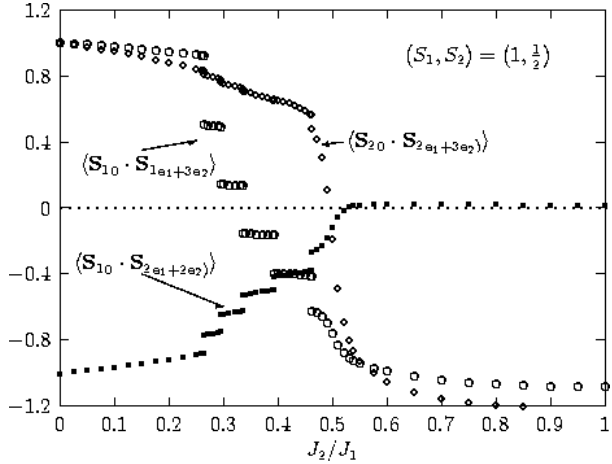,width=8cm}
\vspace{0.3cm}
\narrowtext
\caption{Spin-spin correlators vs. $J_2/J_1$ in the
system with $N=20$  spins  (ED results). The presented functions are
scaled  by their absolute values at $J_2=0$.
}
\label{ss}
\end{figure}
%%%%%%%%%%%%%%%%%%%%%%%%%%%%%%%%%%%%%%%%%%%%%%%

The ED data for different spin-spin correlations (Fig.~\ref{ss}) give
some additional  evidence  in favor of  the suggested phase diagram.
Here the phase transition point $\alpha_{c2}$  
can be indicated as a point where  the mixed-spin correlations 
(being ferromagnetic in the N phase)  change their sign. 
We observe  that close to the point $\alpha_{c2}$
the spin-spin correlations among the $S_2$  spins  are small  and also 
change their sign. On the other hand,  the  correlations among the $S_1$  
spins  remain relatively large  (and antiferromagnetic, as in the N phase) 
down to the point $\alpha^*\approx 0.46$, where the singlet ground state  disappears.  
Thus, there are some indications that the established
quantum spin phase ($\alpha^*< \alpha < \alpha_{c2}$) 
has the symmetries of the canonical  N\'eel state composed entirely of $S_1$ spins
(i.e., $m_1\neq 0$, $m_2=0$). However, due to the finite-size effects in the ED data
we can not exclude the spin-liquid state ($m_1=0$, $m_2=0$) as a possible
ground state in the above region.
Further understanding of the proposed phase diagram 
may be achieved by studying the  low-lying   excitations in these 
magnetic phases. 
%%%%%%%%%%%%%%%%%%%%%%%%%%%%%%%%%%
\subsection{Low-energy excitations}
%%%%%%%%%%%%%%%%%%%%%%%%%%%%%%%%%%
In the F  phase    $B_{1\bf k}= B_{2\bf k}= D_{\bf k}=0$  and  
the dispersion relations in Eq. (\ref{sp}) are simplified to     
\begin{equation}\label{sp1} 
\omega_{1,2}^{(F)}=\sqrt{A_{\bf k}^2
-C_{\bf k}^2}\pm\frac{1}{2}( A_{1\bf k}-A_{2\bf k})\, ,
\end{equation} 
where  
$A_{1\bf k}=4J_1S_2-4J_2S_1(1-\nu_{\bf k})$, 
$A_{2\bf k}=4J_1S_1-4J_2S_2(1-\nu_{\bf k})$, 
$C_{\bf k}=4J_1\sqrt{S_1S_2}\gamma_{\bf k}$,
 $\nu_{\bf k}=\cos k_1\cos k_2$, and  $\gamma_{\bf k}=(\cos k_1+\cos k_2)/2$.
The low-energy physics is controlled 
by the acoustic  branch  $\omega^{(F)}_1$.    
For small $|{\bf k}|$  $\omega^{(F)}_1$  takes the form  
\begin{equation}\label{sp11}
\omega^{(F)}_1=\frac{\rho_s^{(F)}}{m_0}k^2+{\cal O}(k^4)\, ,
\end{equation}
where $\rho_s^{(F)}=J_1S_1S_2-J_2(S_1^2+S_2^2)$
 is the spin-stiffness constant of the ferrimagnetic state
and $m_0=(S_1-S_2)/a^2$ ($a=\sqrt{2}$) is the magnetization density. 
$\rho_s^{(F)}$ remains finite at  the F-C transition
point $\alpha_{c1}$.  The F-C phase transition is 
connected with a  softening of the 
acoustic branch $\omega^{(F)}_1$
at   the corners of the  PBZ. For instance, close
to  ${\bf k}=(0,\pi)$   the magnon  spectrum  reads
$$
\omega^{(F)}_1=8J_1S_1(\alpha_{c1}-\alpha)+4J_1S_1\left[
k_1^2+(\pi-k_2)^2\right]\, .
$$
On the other hand, the optical branch 
$\omega^{(F)}_2$  remains  stable at $\alpha_{c1}$ 
and does not play important  role in  the quantum 
F-C phase transition.
%%%%%%%%%%%%%%%%
\begin{figure}
\centering\epsfig{file=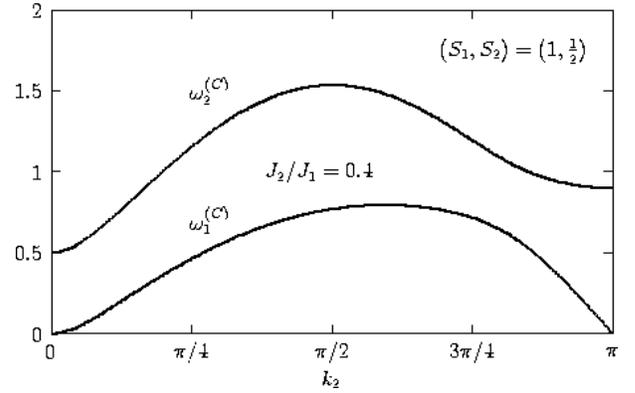,width=8cm}
\vspace{0.3cm}
\narrowtext
\caption{
Spin-wave spectrum  in the C phase 
along the ${\bf e}_2$  direction of the PBZ
($k_1=0$, $J_1\equiv 1$).
}
\label{ec}
\end{figure}
%%%%%%%%%%%%%%%%%%%%%%%

The   excitation spectrum  in the C phase is
described by the general expression (\ref{sp}),  where 
$A_{1\bf k}=4J_1S_2t+4J_2S_1[1-t^2(2-\nu_{\bf k})]$,
$A_{2\bf k}=4J_1S_1t-4J_2S_2(1-\nu_{\bf k})$,
$B_{1\bf k}=-4J_2S_1(1-t^2)\nu_{\bf k}$, $B_{2\bf k}=0$,
$C_{\bf k}=2J_1\sqrt{S_1S_2}(1+t)\gamma_{\bf k}$,
$D_{\bf k}=-2J_1\sqrt{S_1S_2}(1-t)\gamma_{\bf k}$,
and $t^{-1}=2\sigma\alpha$  (see Fig.~\ref{ec}). 
As may  be expected, in addition to the quadratic spin-wave excitations, 
Eq.~(\ref{sp11}),  close to the zone corners there appear linear Goldstone
excitations  connected to the spontaneously 
broken $U(1)$ symmetry in the C  phase. 
For example, in the vicinity of ${\bf k}=(0,\pi)$ the spectrum takes the
form
$$
\omega^{(C)}_{1}=v\sqrt{k_1^2+(\pi-k_2)^2}\, ,\hspace{0.3cm} 
v=4J_1S_1\sqrt{\alpha^2-\alpha_{c1}^2}\, .
$$ 
Note that the spin-wave velocity $v$ vanishes at the F-C transition point
$\alpha_{c1}$.  
%%%%%%%%%%%%%%%
\begin{figure}
\centering\epsfig{file=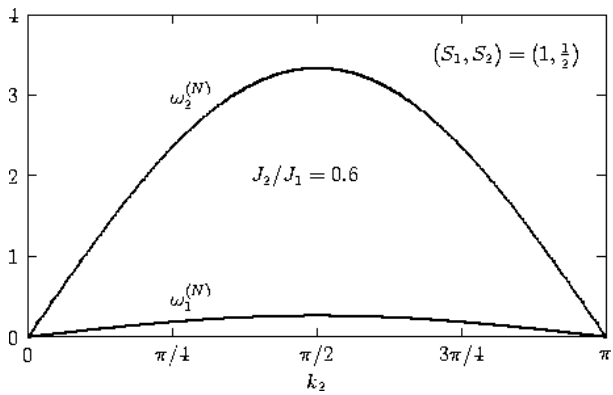,width=8cm}
\vspace{0.3cm}
\narrowtext
\caption{
Spin-wave spectrum  in the N phase 
along the ${\bf e}_2$  direction  of the PBZ
($k_1=0$, $J_1\equiv 1$).
}
\label{en}
\end{figure}
%%%%%%%%%%%%%%%%%%%%%%%%%%%%%

Finally, the spin-wave  spectrum in the  
mixed-spin collinear  phase
is given by the general relation (\ref{sp}), where 
$A_{1\bf k}=4J_2S_1$, $A_{2\bf k}=4J_2S_2$,
$B_{1\bf k}=4J_2S_1\nu_{\bf k}$,
$B_{2\bf k}=4J_2S_2\nu_{\bf k}$,
$C_{\bf k}=2J_1\sqrt{S_1S_2}\cos k_2$,  and
$D_{\bf k}=2J_1\sqrt{S_1S_2}\cos k_1$ (see Fig.~\ref{en}).  
As compared to the antiferromagnetic $J_1-J_2$ model,
the excitation spectrum in  the mixed-spin  collinear phase 
shows some  important peculiarities  which specify  the instability
at the classical transition point $\alpha =0.5$.  First, 
in the mixed-spin system the degeneracy of
the spin-wave branches  $\omega_1^{(N)}$ and $\omega_2^{(N)}$ 
is completely removed. Second, the spin-wave 
excitations described by $\omega_2^{(N)}$ formally remain  
stable down to the point  $\alpha=\sqrt{\sigma}/(\sigma +1)$  
[$=\sqrt{2}/3\approx 0.47$ for the $(1,\frac{1}{2})$ system].
The phase transition at $\alpha_{c2}$ is entirely driven by    
instabilities in the lower $\omega_1^{(N)}$  excitation branch.  
At the classical spin-flop transition point ($\alpha =0.5$),  
the branch $\omega_1^{(N)}$ exhibits two full lines 
of zero modes  (i.e., $k_1=0$ and $k_2=0$)  in the PBZ. 
The soft lines are  connected with {\em ferromagnetic} fluctuations  
in the ${\bf e}_1$ and  ${\bf e}_2$ directions on the square lattice,
so that we can  expect that quantum fluctuations  
do not change the position of the classical transition 
point  (in accord with the data
presented in Fig.~\ref{mm}).\cite{comm2} 
%%%%%%%%%%%%%%%%%%%%%%%%%%%
\section{Conclusions}
%%%%%%%%%%%%%%%%%%%%%%%%%%%
In  conclusion,  we have studied the quantum spin phases 
in the mixed-spin $J_1-J_2$ Heisenberg model defined on a square lattice.
The reported results support a phase diagram  
with an additional quantum spin  phase  close to the  classical spin-flop
transition. In particular, in the extreme quantum case $(1,\frac{1}{2})$, 
we have found  indications of a quantum spin phase 
(in the region $0.46< \alpha < 0.5$) which is characterized by
magnetically disordered $S_2$ spins and   
N\'eel long-range ordered $S_1$ spins. The latter suggestion is based 
(i) on the analysis of the low-lying excitations 
showing that the mixed-spin collinear state survives at most down to the classical 
transition point $\alpha=0.5$ and (ii) on the SWT and ED results
for the sublattice magnetization in the C phase $m_2$: both methods predict
that $m_2$ vanishes at a  point ($\alpha\approx 0.46$)  preceeding the
classical transition point $\alpha =0.5$. 
However, due to strong finite-size effects
in the ED data and the qualitative character of the SWT analysis,
we can not exclude the spin-liquid state as a possible 
ground state in the above region. Studies with other analytical and
numerical methods would be necessary to characterize the true 
ground state in the above region. 
%%%%%%%%%%%%%%%%%%%%%%%
\acknowledgments
%%%%%%%%%%%%%%%%%%%%%%%%
This work was partially  supported by the Bulgarian Science 
Foundation (Grant No. 817/98)
and the Deutsche Forschungsgemeinschaft (Grants No. Ri615/7-1 and 430 BUL 17/2/02). 
%%%%%%%%%%%%%%%%%%%%%%%%%%%%%%%
 
\end{multicols}

\begin{references}
\bibitem[*]{ivanov} Permanent address: Institute  of Solid  State Physics, 
Bulgarian Academy of Sciences,
Tzarigradsko chaussee 72, 1784 Sofia, Bulgaria.
\bibitem{sarma}  S. Das Sarma, S. Sachdev, and L. Zheng, Phys. Rev. B
{\bf 58}, 4672 (1998),   and references therein.
\bibitem{yang} At filling factor $\nu =2$ the system can be mapped
onto the spin-$\frac{1}{2}$  easy-plane Heisenberg ferromagnetic model 
subject to a magnetic field along the $z$ direction:
K. Yang,  Phys. Rev. B {\bf 60}, 15 578 (1999).
\bibitem{troyer} Y. Matsushita, M. Gelfand, and C. Ishii,
J. Phys. Soc. Jpn. {\bf 66}, 3648 (1997);
M. Troyer and S. Sachdev, Phys. Rev. Lett. {\bf 81}, 5418 (1998).
\bibitem{sachdev} S. Sachdev, Z. Phys. B {\bf  94}, 469 (1994);
S. Sachdev and T. Senthil, Ann. Phys. (NY) {\bf 251}, 76 (1996). 
\bibitem{sushkov}  L. Capriotti, F. Becca, A. Parola, and
S. Sorella, Phys. Rev. Lett. {\bf 87}, 097201 (2001);
O.P. Sushkov, J. Oitmaa, and Z. Weinhong,
Phys. Rev. B {\bf 63}, 104420 (2001).
\bibitem{kahn} O. Kahn, {\em Molecular magnetism} (VCH, New York, 1993).
\bibitem{comm3}  The classical N phase is degenerate in respect to 
global rotations of the $S_1$  (or $S_2$) subsystem. Quantum fluctuations 
choose one of the collinear configurations corresponding 
to ferromagnetic spin arrangements along, say,  the
${\bf e}_1$  or the   ${\bf e}_2$ directions on  the square lattice. 
\bibitem{farnell}  Below we present extrapolated resuls  obtained 
with  the  SUB$m$-$m$ approximation scheme ($m=2,4$,
and $6$ in the F and N phases;
$m=3,4$, and $5$  in the C phase). For a recent elaborate
description of the CCM method, see D.J.J. Farnell,
K.A. Gernoth, and R.F. Bishop,
Phys. Rev. B {\bf 64}, 172409 (2001).
\bibitem{wald} Such a method to obtain the pitch angle 
has previously been used by Ch. Waldtmann et al., 
Phys. Rev. B {\bf 62}, 9472 (2000).
\bibitem{comm0}
Since in the SWT for fixed $S_2=\frac{1}{2}$ the position
of the point $\alpha^*$ ($< 0.5$) monotonically changes towards
larger $\alpha$ for larger $S_1$, it is natural to
expect that the instability at $\alpha^* < 0.5$ exists
in the extreme quantum case $S_1=1$ as well.
\bibitem{comm1}   The region beyond $\alpha^*$ can  be studied
in the framework of  the  modified spin-wave theories: 
M. Takahashi, Phys. Rev. B {\bf 40}, 2494 (1989);
J.E Hirsch and S. Tang, Phys. Rev. B {\bf 40}, 4769 (1989);
Q.F Zhong and S. Sorella, Europhys. Lett. {\bf 21}, 629 (1993).
\bibitem{comm2}  Note that in the classical mixed-spin collinear 
phase the chain configurations  
in the ${\bf e}_2$  direction are characterized with  
a finite ferromagnetic moment per cell. As a result,
the soft line $k_1=0$ in the mixed-spin state  is connected with  
ferromagnetic spin fluctuation. On the other hand,
the above ferromagnetic moment is zero in
the uniform-spin ($S_1=S_2$) collinear phase,
so that the soft line $k_1=0$  is
connected with antiferromagnetic fluctuations.
This explains the change of the position
of the $\alpha_{c2}$ point  ($\alpha_{c2}\approx 0.6$) found in the
spin-$\frac{1}{2}$ $J_1-J_2$ Heisenberg model.
\end{references}
\end{document}